# COMPARATIVE ANALYISIS ON SOME POSSIBLE PARTNERSHIP SCHEMES OF GLOBAL IP EXCHANGE PROVIDERS

David Gunawan[1] and Karno Budiono[2]

Research & Development of Infrastructure,
Telkom Indonesia Research & Development Centre, Bandung, Indonesia

*ABSTRACT*

*IPX (IP eXchange) is GSMA's proposal for IP interconnection model which supports multi services to offer end-to-end QoS, security, interoperability, SLAs through a dedicated connection. It provides a commercial and technical solution to manage IP traffic and follows the GSMA's 4 key IP interworking principle such as openness, quality, cascading payments, and efficient connectivity. In order to get global IPX reachability, it is possible for an IPX provider to build partnership with other global IPX providers in business and network configuration. There are some possible partnership schemes between IPX providers such as peering mode, semi-hosted mode, full-hosted mode, or combination between these modes. The implementation of the schemes will be case-by-case basis with some considerations based on (but not limited to) IPX Provider's network asset & coverage, services & features offer, commercial offer, and customers. For an IPX provider to become competitive in IPX business and become a global IPX hub, some value added should be considered such as cost efficiency and great network performance. To achieve it, an IPX provider could implement some strategies such as build network sinergy between them and partners to develop IPX Service as single offering, offer their customers with bundled access network and services. An IPX provider should also consider their existing customer-based that can be a benefit to their bargaining position to other potential IPX provider partners to determine price and business scheme for partnership.*

## 1. INTRODUCTION

Nowadays, as international telecommunication business increases in means of service types, traffic, and operator revenues, then IPX become one of telecommunication operator's option as an interconnection model that support multi services for their customers. A number of services such as roaming data 2G/ 3G/ LTE, roaming signaling 2G/ 3G/ LTE, SMS/ MMS interworking, RIM connection, WiFi roaming, bilateral IPX services, Voice over IPX, HD Voices, and RCS roaming can be delivered through IPX connection. Based on GSMA definition, IPX is a telecommunications interconnection model for the exchange of IP-based traffic between customers of separate mobile and fixed operators as well as other types of service provider (such as ISP), via IP based network-to-network interface. In the interconnection context, IPX is used to mean an interconnection at the service level (not at the network level). It also refers to the collection of all the interconnected IPX provider's networks, a subset of the inter-service provider IP backbone. The IPX network includes inter-service provider IP backbone which comprises the interconnected networks of various IPX providers. An IPX provider is a provider that offers IPX services, meanwhile a service provider is a mobile, fixed operator, or other types of operator connecting to inter-service provider IP backbone for roaming and/or interworking purposes.
As the next generation interconnect solution, IPX have a number characteristics, such as:

- Openess, means that any potential players in the delivery of IP Services (MNOs, FNOs, Carriers and ISPs) has the freedom of choice to be involved
- Quality, means reliable & secure delivery of services in conformance to agreed QoS levels ensures benefits for all player and end users
- Cascade payments, means parties who meet their mutual obligations in the value chain will receive a fair commercial return
- Efficient connectivity, means IPX is a gateway to managed IP network- managing data flow and commercial information and providing the benefits of multilateral connectivity to all players

Generally, a service provider have two possibilities to interconnect with other service providers either by establishing an IPX connection via IPX providers (or GRX providers if only for the GRX service) or using direct connection with other service providers with leased lines, internet using IPSec protocol, or VPN connectivity. Interconnection using IPX is shown in Figure 1, which service provider A uses IPX provider X to interconnect with service provider B and C. IPX provider X have direct connection with service provider A and B as on-net subscriber means that it will be no problem to have interconnection between service provider A and B since they belongs to same IPX provider. However, IPX provider X should cooperate with IPX provider Y in order to service provider A possible to interconnect with service provider C since IPX provider X doesn't have direct cooperation with service provider C. This is the basic need for IPX providers cooperation.

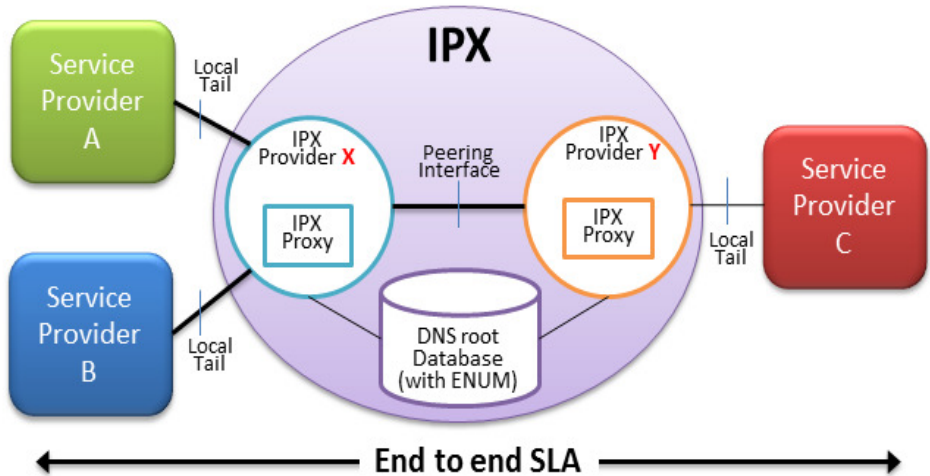

Figure 1. IPX basic network configuration

The main background of cooperation between IPX providers because the difficulties for one IPX provider to have a global and direct connectivity to all service providers in the world since the will takes time and strong effort in business and network infrastructure aspect. One of ideas for an IPX providers to solve that problem is through cooperation and partnership with other IPX providers. The goal of this paper is to analyze some possible partnership schemes of global IPX providers.

The remainder of the paper is structured as follows: In section 2, we analyze current condition of IPX providers from technical and business aspects, which include IPX capability, development drivers, barriers, potential business models and revenue stream from IPX. In section 3, we continue with some IPX interconnection model which consists of IPX bilateral transport only, IPX service transit, and IPX multilateral hub services. In section 4, we explain some possible IPX partnership models between IPX providers such as normal IPX peering, semi hosted, and full hosted. In section 5, we conclude the paper with some recommendations to choose the most suitable partnership models for IPX providers.

## 2. ANALYSIS OF CURRENT CONDITION OF IPX PROVIDERS

IPX basically is a technology evolution of GRX therefore the providers and market itself are already quite mature. An IPX provider is possible to offer multiple type of telecommunication services with single IP network connection and end-to-end network performance guarantees. In the other hand, network elements of IPX still similar with GRX but with addition of Diameter router to accommodate LTE roaming service. The emerging market for IPX is LTE-based roaming services (signalling, voice, and data). However, the OTT (Over The Top) providers markets still wait for the strong drivers to use IPX. Even, the bigger bandwidth in customer side make OTT can still use public internet network as happened today. IPX is also able to support a number of GRX services such as MMS interworking and WLAN (authentication) data roaming, as well as diagnostic protocols, for example ICMP (Ping), connectivity between any types of service providers, end-to-end QoS for roaming and interworking, and any IP services on a bilateral basis with end-to-end QoS and interconnect charging.

Some drivers for IPX development come from both IPX providers and service providers such as from technology background to migrate circuit-switched services to IP, LTE interoperability, LTE roaming, and some new retail services (HD voice, high quality video services). From business background, IPX bring opportunities in some aspects such as introducing new revenue-generating services, increasing quality, the cost and operational advantages of the hub model for service interconnect, and could drive out cost by combining multiple services over a single connection.

Despite some drivers listed above, a number of IPX providers and service providers also consider some barriers to develop IPX. From IPX provider's point of view, they will face organizational barriers including operational splits between voice and data, fixed and mobile, commercial and network department, lack of critical mass means that many were not prepared to migrate of there only a few partners using an IPX, lack of LTE network and no visible time line for LTE launch, and services pricing issues that their potential customers didn't get detail pricing information clearly from them. From service provider's point of view, some barriers to develop IPX are lack of IPX understanding that many of them still not convince with IPX capability because of minimum IPX knowledge, uncertainty about the ability IPX to fix interoperability problems that a few IPX providers fail to make adequate information available. The barriers could also come from regulation and infrastructure perspective such as license of international service providers, restriction/ outright ban on VoIP, and infrastructure/ geographical barrier that lack of international IP connectivity/ capacity in many emerging countries.

There are a number of possible business model and role strategies for IPX such as IPX network operator that build PoPs in key geographic markets connected by an MPLS networks, IPX platform provider that lease network services and focus on providing interoperability platforms, white label reseller that focusing on selling access to third party networks and platforms – possibly on a white label basis, VAS Provider that focusing on value added services for IPX providers to resell or build communities of application providers for them, Voice IPX specialist

International Journal of Computer Networks & Communications (IJCNC) Vol.6, No.2, March 2014

that ignore the data service market in the short term and focus on VoIPX only, Regional gateway that seek to build a strong regional IPX network and service offering, and non-IPX player that stay away form IPX altogether and focus on providing high quality voice, GRX, and signaling services, building on what carrier already does now.

IPX networks are being considered for, or used as, a platform for the delivery of a variety of new international or roaming services. The services which scored highest for both 'currently using' or 'already plan to use IPX' were GRX and enhanced GRX, roaming signalling for LTE and legacy services, SMS and MMS interworking, LTE voice, LTE data roaming, and content services. Other services such as HD voice and TDM/ VoIP interoperability also possible to be implemented using IPX.

One example of IPX potential revenue streams come from managed access services that not only offer services, but also for access connection, another business model typically purchase of connection, port, and capacity. Other revenue streams are from roaming data transit services (CRX, GRX, and LTE roaming), roaming IP-based LTE voice (VoLTE) transit, roaming signalling (transport for 2G, 3G, roaming signalling, and LTE signalling), roaming messaging (roaming SMS and MMS), settlement and clearing (data, financial clearing, and settlement naturally as VAS for IPX providers), traffic steering for a variety of guises including traffic redirection using mobile number portability database & ENUM database, and analytics that helping IPX customers to improve their service and profitability. One of the example applications are route management and balancing based on QoS, pricing, and knowledge of the number of hops to end points, silent roamer identification and marketing services, fraud management services that enhanced with the use of analytics.

An IPX Provider is also possible to provide NRTRDE (Near Real Time Roaming Data Exchange) services, international voice break-in/ break-out that Provide termination for inbound services on to PSTN or mobile network (break-in), or transit & termination for outbound services, IP Transit with added QoS/ security which includes transit of IP traffic related to cloud services, content, and application. Another example that are already implemented are interconnection of operators' IPX network with RIM data centres to ensure more secure transit of BBM traffic, and hosted application for hosting of managed cloud-based RCS solutions, conferencing solutions, or hosting of enterprise cloud platform (PaaS) that operators can used to serve their end customer with guaranteed QoS assurance. It is also possible to deliver IPX advances telephony such as HD voice and conferencing video calling/ video conferencing (in SD and HD), IPX RCS and rich media by providing interconnection and interoperability for the services, content transcoding and trans—rating that can help operators to deliver content internationally using codecs. Roaming WiFi is one example of popular IPX services which enable mobile operators to take advantage of public WiFi infrastructure in other markets while retaining the ability to monitor customers' usage an to bill customers by using WiFi roaming exchange.

## 3. ANALYSIS ON SOME IPX INTERCONNECT MODELS

In this section, we describe three IPX interconnect models which are possible to be implemented by service providers that are free to choose on a per service basis:

1)  Bilateral Transport Only

In this model, IPX provider provides transport at a guaranteed QoS and each service provider will pay their respective IPX provider costs for transport. The bilateral agreement is between end service providers and any payment of termination charges is a mater for the service providers. A bilateral connection between two service providers (SP-1 & SP-2) using the IPX transport layer with guaranteed QoS end-to-end. In this case, settlement is independent of the IPX domain but connectivity still operates within IPI key business principles. Cascading of responsibilities (such as QoS) applies but not cascading of payments (cascade billing). Each service provider will also pay their respective IPX provider for the transport capacity, potentially depending on the level of QoS provided.

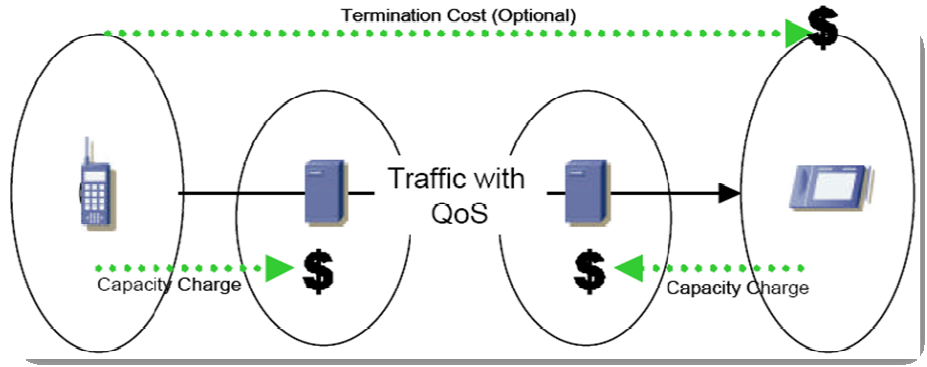

Figure 2. IPX bilateral transport only interconnect model

2) Bilateral Service Transit

The IPX provides QoS-based transport and cascading interconnect payment facilities. This enables an originating service provider to make a single payment to their IPX Provider who passes on a payment on to the next IPX provider in the value chain who pays the final termination charge to the terminating service provider. Within service transit, traffic is transited though IPX providers but prices (termination charges) are agreed bilaterally between service providers and settlement of termination charges can be performed bilaterally between the service providers or via the IPX providers (upon the service provider's choice). Cascade billing (for transport and/or service layer) and other associated facilities provided by the IPX provider (on the service layer) may be applied depending on the service.

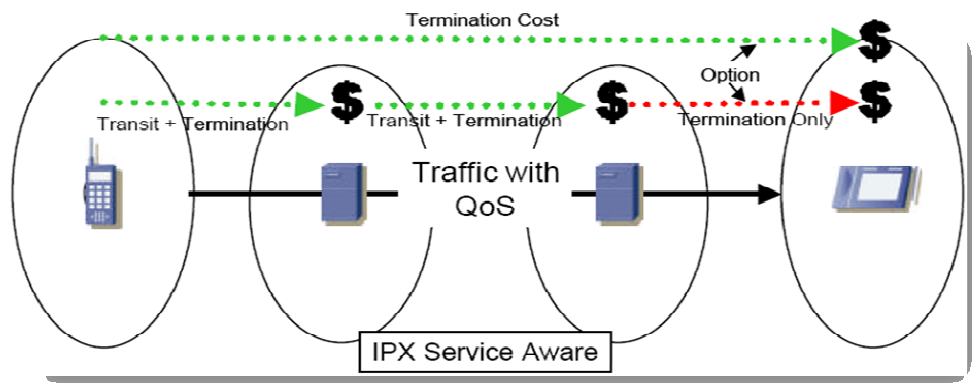

Figure 3. IPX bilateral service transit interconnect model

International Journal of Computer Networks & Communications (IJCNC) Vol.6, No.2, March 2014

A bilateral connection between two service providers (SP-1 & SP-2) using the IPX service layer and the IPX transport layer with guaranteed QoS end-to-end. Within service transit, traffic is transited though IPX providers but prices (termination charges) are agreed bilaterally between service providers and settlement of termination charges can be performed bilaterally between the service providers or via the IPX providers (upon the Service Provider's choice).

Cascade billing (for transport and/or service layer) and other associated facilities provided by the IPX Provider (on the Service layer) may be applied depending on the service. Therefore, through service transit, the following connections can be implemented:

- Bilateral connectivity with routing performed within the IPX domain and within IPI key business principles but settlement of termination charges performed bilaterally between the ending parties.
- Bilateral connectivity with both routing and settlement of termination charges performed under the IPX Domain and within IPI key business principles.

The transit fee owed to the IPX Providers is always cascaded. Cascading of responsibilities and payments fully apply (on both IPX transport layer and IPX service layer).

3) Multilateral Hub Service

IPX provides QoS transport and cascading interconnect payments to a number of interconnect partners via a single agreement between the service provider and IPX. This "one-to-many" mode is operationally highly efficient for the service provider. Charging transparency is a requirement on both IPXs and service providers in this multilateral connection using hub functionality. Hubbing or multilateral connectivity is where traffic is routed from one service provider to tens/ hundreds of destinations/ interworking partners through a single agreement but the cascading of responsibilities applies. Cascading of payments may be applied depending on the service (on both IPX transport layer and IPX service layer).

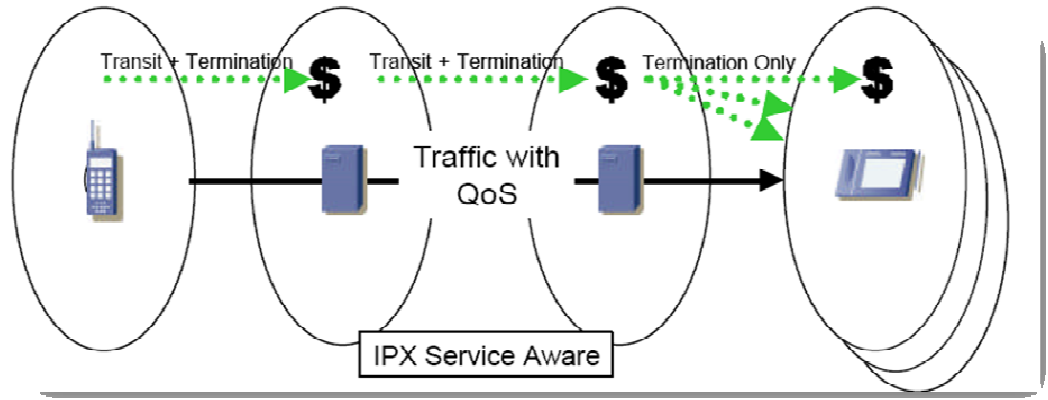

Figure 4. IPX multilateral hub service interconnect model

The deployment scenarios are possible to be implemented using two alternatives. The first option is through direct investment. The benefit of this option are total operational control, access to new markets, flexibility in choosing geographic location. But, this option also have some drawbacks such as longer time to market, requires capital commitment, and high risk. IPX providers should consider some strategy before implement this option such as investing on data roaming services as the first stages of IPX deployment and initiate peering partnership with other potential IPX Provider.

The second option is through collaboration. The benefit of this option are sharing of risk, fast roll out, access to existing infrastructure, and geographic network. However, this option have some drawbacks in smaller profit margin, dependant on partner's strategy. The implementation strategies could be collaboration with leading IPX provider to resell (white label) under its own brand that enable immediate access to IPX services range and coverage, or collaborate without white label scheme.

Some existing IPX providers' background are experienced GRX (GPRS Roaming eXchange) providers and IPX implementation is executed with strategy to add IPX capability over their existing network. It means that currently all GRX operators are IPX-ready and they are in progress in partnership stages to extend their coverage area and potential customers. The partnership itself is already built from GRX that then developed to IPX. A number of customers and coverage areas become main considerations to choose partnership model, whether based on peering and/ or transit.

## 4. ANALYSIS OF POSSIBLE IPX PARTNERSHIP SCHEMES BETWEEN GLOBAL IPX PROVIDERS

The main idea of IPX partnership between IPX providers come from the limitation of one IPX provider to have global coverage to all their potential customers all over the world. The generic configuration for IPX partnership is shown in figure 5.

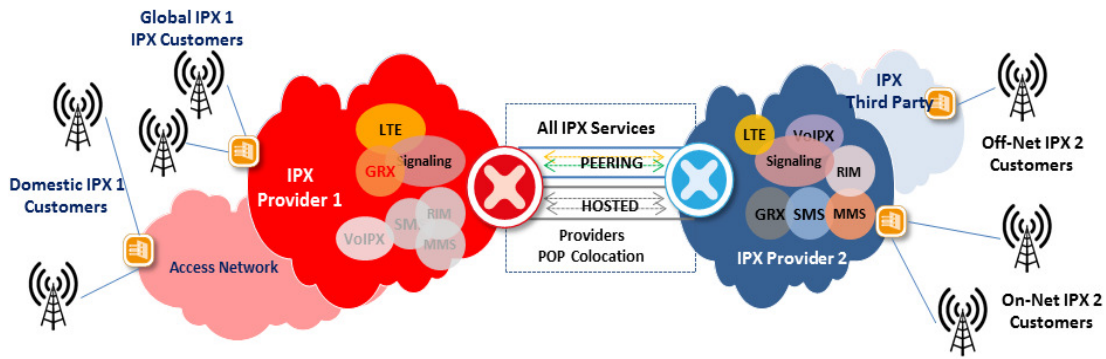

Figure 5. IPX providers partnership generic configuration

Some key points regarding IPX partnership model shown in Figure are:

- Partnership model could be peering and/ or hosted based on service
- Network responsibilities L1/ L2 network peering at POP location with bandwidth capacity and QoS will be based on further agreement and requirements between partners
- IPX services implementation could be implemented gradually based on agreements between partners
- Business scheme and charging could be based on type of IPX customers (on-net, off-net, location) and traffic volumes. In most cases, all on-net customers will be opened and charged based on traffic activities

105

From above generic configuration, there are at least 3 (three) possible IPX patnerships could be implemented comprises of normal IPX peering, semi hosted, and full hosted partnership schemes.

### 4.1. Normal IPX Peering

The normal partnership between IPX providers is based on IPX peering shown in Figure 6.

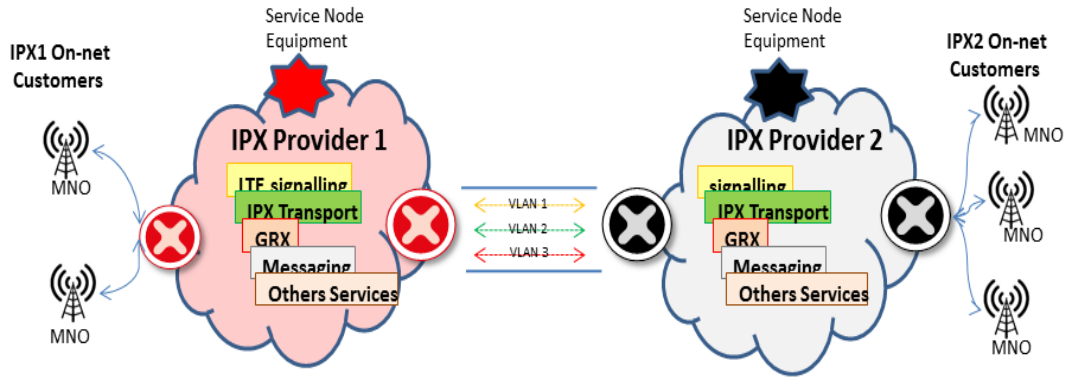

Figure 6. Normal IPX peering between 2 (two) IPX providers

In standard peering model between IPX providers defines NNI (Network to Network Interconnection) and the access is limited to on-net (direct) partner's IPX customers. In this model, VLAN should be separated based on service, and the traffic and charging will be consolidated for all MNOs per-service based. The reporting also should be based on service and there is no dedicated reporting per MNO. In normally commercial model, it is possible to add instalation fee and monthly fee parameters based on type of services and number of destinations between IPX providers.

The main advantage for this partnership model is both partners already have independent service node elements and system, and they will have same position level and can reach or access IPX partner's on-net customers. The challenges of this model are each IPX provider need to peer with more than one IPX providers to get global reach since majority of IPX providers will not open their off-net destinations. Some cases will be occured when a larger IPX provider peer with the smaller one means that the smaller IPX provider need to pay to the bigger one. Other notes for this partnership model are IPX network will be separated per-VLAN-based per-service and consolidated traffic & reporting per-service will be in IPX provider's level.

### 4.2. Semi Hosted IPX Partnership

The semi hosted IPX partnership is shown in Figure 7.

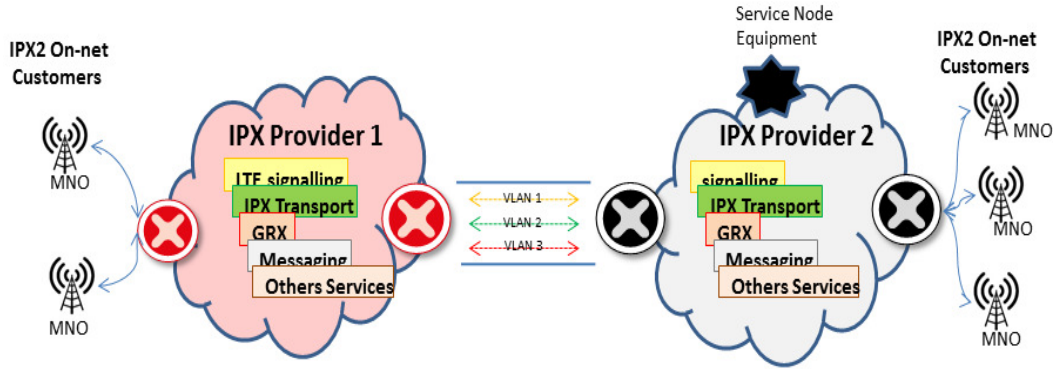

Figure 7. Semi hosted IPX partnership between 2 (two) IPX providers

In this partnership model, IPX provider 1 (the left one) doesn't need to invest their own service node equipments and IPX system since they will use elements and systems from IPX provider 2 (the right one). This model can be used as starting point and short-term scenario for a new established IPX provider that already network infrastructure and prospective customers and they want to deliver IPX services instantly to their customers without build their own IPX system.

IPX provider 1 is possible to have access to IPX provider 2's complete IPX coverage (on-net and off-net) and it could minimize the possibility to have partnership with other IPX providers. IPX Provider 2 will manage IPX service node elements for all or specific services. In the other hand, IPX provider 1 will provide CPEs in customer's side and access network from customer to IPX provider 2's service node. It also should separate VLAN per service and consolidated traffic for all MNOs by service. The consolidated reporting will base on service and no dedicated reporting per MNO. In normally commercial model, this partnership scheme is often included installation fee and monthly fee per MNO (based on bandwidth size or per message)

However, this partnership scheme will lead to exclusive partnership between both IPX providers and IPX provider 1 will depend on capability and coverage of IPX provider 2's. There are also possibility for some business issues such as price competitiveness and margin share between partners. Same with the first model, the IPX network will be separated per-VLAN-based per-service and consolidated traffic & reporting per-service in IPX provider's level.

### 4.3. Full Hosted IPX Partnership

The full hosted IPX partnership is shown in Figure 8.

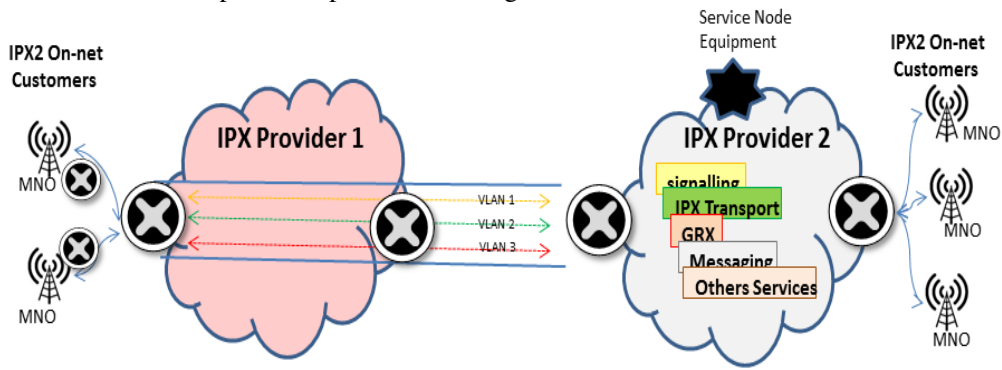

Figure 8. Full hosted IPX partnership between 2 (two) IPX providers

International Journal of Computer Networks & Communications (IJCNC) Vol.6, No.2, March 2014

This partnership scheme almost similar with semi hosted partnership explained above that IPX provider 1 possible to access IPX provider 2's on-net and off-net customers and they don't need to invest on IPX service node equipment. The main difference between them is in full hosted partnership scheme, IPX providers 1's position and main task is to market IPX provider 2's IPX service since all IPX network infrastructure will be provided by IPX provider 2. However, there is a possibility for IPX Provider 1 to re-brand the IPX services using their own brand. IPX provider 1 can apply a direct price mark up for reselling.

By considering existing services, infrastructure, potential partnership, market, and implementation time, an IPX operator can have different types of partnership model, i.e. some IPX providers is come from voice services and signalling providers, they can migrate voice traffic to IPX-based. For example, an IPX provider that already have existing strong customers and partners can attract other IPX providers to peer with them. The implementation model could be started by migrating traffic from non-IPX to IPX environment without changing the existing business model. Some new services such as LTE (signaling, data, and voice) and diameter could become main drivers to do partnership between IPX providers since LTE is a green-field service that currently in initiation stages, with some IPX providers willing to have peering and do trial with other IPX providers.

For other services such as GRX, SMS/MMS, and RIM can be implemented using aggregation business model since GRX and SMS/MMS are mature services then the performance improvement resulted in IPX environment is still can not be a major driver for service providers to move to IPX. In some cases, a number of GRX providers offer aggregator partnership model, with consideration in implementation simplicity. A non-GRX provider is possible to facilitate access network from their existing and domestic customers to their IPX service node provider partner. By having partnership with them, they will become GRX/SMS/MMS/RIM hubber.

**4.4. Consideration to choose IPX partnership scheme**

The IPX partnership between IPX providers come from business and technical perspective. From business perspective, it could be seen 3 (three) issues, such as:

- Peering scheme (IPX transport with services) with equivalent IPX providers with a number of on-net customers and coverage areas as main consideration to build partnership
- Transit scheme with non-equivalent IPX providers
- For peering scheme, usually only include on-net customers of peering partners. Therefore, peering will need more than 1 (one) IPX providers to reach global coverage. Based on experience from some IPX providers, they could build peering partnership with more than 5 (five) other IPX providers.

From technical perspective, there are several issues regarding IPX partnership:

- To maintain network performance, majority of IPX providers limit their end-to-end IPX customers to maximum 2 IPX providers (2 hops).
- Several IPX providers stated that peering partnership could be a challenging task for service interoperability.
- Although IPX offer single IP private connection for multi services, however the reporting mechanism is often done partially.

## 5. CONCLUSIONS

There are a number of possible partnership schemes can be implemented between IPX Providers such as peering, semi hosted, full hosted, or combination between with service-based implementation. However, deciding the best partnership scheme, IPX providers should consider some factors, related to (but not limited to) IPX providers' network asset, coverage, and ownership, IPX services and features offering whether they offer a part or full IPX-based services, and their support of tools and data analytic, financial data clearing, ENUM, CDN, or other IPX-related services. In the other hand, a number of a IPX Provider's on-net and off-net customers should become one main considerations, beside business scheme and pricing offer.
For an IPX provider to become competitive in IPX business and become a global IPX hubber, they should able to give value added to customers, such as cost efficiency and great network performance. To achieve it, an IPX provider could implement some strategies such as build network sinergy between them and partners to develop IPX Service as single offering, offer their customers with bundled access network and IPX Service with cheaper price than competitors. An IPX provider should also consider their existing customer-based that can be a benefit to their bargaining position to other potential IPX provider partners to determine price and business scheme for partnership.

**Authors**

1. David Gunawan

**Position:** Engineer 2 Sofswitch System – Telkom R&D Center
**Projects Experiences:**
- Project Manager for Prototype Web Portal User Provisioning IMS  - 2013
- Project Manager for Research of Integration IMS with EPC – 2012
- Project Manager for Telkom Softswitch Standard Update – 2011

**International Activities:**
- ITU (International Telecommunication Union) Academy Member, 2012-now

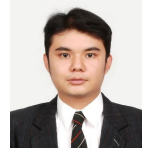

2. Karno Budiono

**Position:** Researcher Packet IP Based Service Node – Telkom R&D Center
**Projects Experiences:**
- Project Manager Assessment Metro Network to Support IDN 2015

**International Activities:**
- Representative in  the 3nd Plenary Meeting Tropic Project, Rome,2013
- Speaker in  the MPLS World Congress, Paris, 2010
- Telkom representatives in MEF Quarterly Meeting, Washington DC, 2009

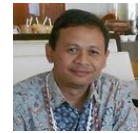